\documentstyle[12pt]{article}

\newcommand{\be}{\begin{equation}}
\newcommand{\ee}{\end{equation}}
\newcommand{\ba}{\begin{eqnarray}}
\newcommand{\ea}{\end{eqnarray}}
\begin{document}
\def\pct#1{\input epsf \centerline{ \epsfbox{#1.eps}}}

\begin{titlepage}
\hbox{\hskip 12cm ROM2F-96/14  \hfil}
\hbox{\hskip 12cm \today \hfil}
\vskip 1.5cm
\begin{center} 
{\Large  \bf  Open \ Superstrings\footnote{Talk presented at the ``IV
Italian-Korean Meeting on Relativistic Astrophysics'', Rome~-~Gran
Sasso~-~Pescara, July 9-15, 1995.}}
 
\vspace{1.8cm}
 
{\large \large Gianfranco \ \  Pradisi}

\vspace{0.8cm}

{\sl Dipartimento di Fisica\\
Universit{\`a} di Roma \ ``Tor Vergata'' \\
I.N.F.N.\ - \ Sezione di Roma \ ``Tor Vergata'' \\
Via della Ricerca Scientifica, 1 \ \
00133 \ Roma \ \ ITALY}
\vspace{0.5cm}
\end{center}
\vspace{3cm}
\abstract{We review the basic principles of the construction of open
and unoriented superstring models and analyze some representative examples.} 
\vfill
\end{titlepage}
\addtolength{\baselineskip}{0.3\baselineskip} 

\vskip 24pt
\begin{flushleft}
{\large \bf 1. \ Introduction}
\end{flushleft}

(Super)String theories \cite{gsw} have emerged in the last 20 years as the only
candidate to describe in a unified scheme all the fundamental interactions,
including gravity.   Although closed oriented models are often 
thought to be more
elegant and promising, we would like to convince the reader that models of
closed and open unoriented (super)strings (called for simplicity open) own the
same level of consistency.  This is also suggested from the recent ideas on
string dualities, according which all (super)string theories, being connected
in a non-perturbative way, are different manifestations of an underlying,
yet unknown, entity \cite{htheory}.  Perturbatively, open (super)strings and,
say, heterotic strings \cite{het} look very different, but there are evidences
for non-perturbative dualities connecting them \cite{polwit}.  

In this talk we shall review the fundamental aspects of
the construction of open and unoriented (super)string models.  In
section 2, we briefly summarize how to define closed 2-D conformal field
theories (CFT), the basic building blocks of closed (super)strings.  In
section 3 we define CFT on Riemann surfaces with holes and/or crosscaps and
review our algorithm for associating a class of ``open descendants'' to 
left-right symmetric closed CFT using a ``parameter space orbifold''
construction \cite{cargese}.  In particular, we shall be able to describe the
perturbative spectra of models with corresponding internal (Chan-Paton)
symmetry groups and related patterns of symmetry breaking.  Finally, in
section 4, we analyze some significant examples.

\vskip 24pt
\begin{flushleft}
{\large \bf 2. \  Closed CFT and superstrings}
\end{flushleft}

The most important property of 2-D (closed) CFT \cite{bpz} is the factorization
of the stress-energy tensor into chiral components $T(z)$ and
$\bar{T}(\bar{z})$, each a function of the single variable $z$, respectively
$\bar{z}$.  The two chiral sectors of the theory are almost independent and
can be separately solved, provided that, at the end, $\bar{z}$ is identified
with the complex conjugate of $z$.  This implies that the observable algebra
splits into a tensor product of chiral algebras 
\be
{\cal A} \bigotimes {\bar{\cal{A}}}
\label{tenpro}
\ee
each containing the stress-energy tensor and thus the Virasoro algebra.  In
the operator formalism, the decomposition (\ref{tenpro}) corresponds to the
factorization of the Hilbert space as well.  To be precise, conformal fields
are organized in representations of the full symmetry algebra of the model. 
The space of states, however, is the superposition of irreducible
representations (or, better, of superselection sectors):  
\be
{\cal H} \ = \ {\bigoplus}_{{\Lambda} , {\bar{\Lambda}}} \ {\cal
H}_{\Lambda} \otimes {\bar{\cal{H}}}_{\bar{\Lambda}}  \qquad . 
\label{hilbtp}
\ee
The sum in (\ref{hilbtp}) is, in general, over an infinite number of
sectors.  Rational CFT \cite{moosei} are characterized by the fact that the sum
in  (\ref{hilbtp}) contains a finite number of terms.  Bulk conformal fields
are ``intertwining operators'' (not chiral) between different sectors, and can
thus be decomposed in sums (finite in RCFT) of products of {\it chiral vertex
operators} (CVO) \cite{moosei} \cite{cvo} \cite{yassen}
\be
\phi_{k, \bar{k}} \, (z,\bar{z}) \ = \ \sum_{i, \bar{i},
f, \bar{f}} \ {{V_{k}}^f}_i (z) \
{{{\bar V}_{\bar k} \, }^{\bar f}}_{\bar{i}} (\bar z) \
{\alpha^{i  \bar i}}_{ \, f \bar{f}} \qquad .
\label{confields}
\ee
${{V_{k}}^f}_i (z)$ denotes a field in the sector $k$ of conformal
dimension $\Delta_k$ acting on a state $i$ and producing a state $f$, that
can be non trivial only if $f$ is in the fusion of $i$ and $k$.  In the
simplest (diagonal) cases ${\alpha^{i  \bar i}}_{ \, f \bar{f}} = \delta^{i 
\bar i} \delta_{f \bar{f}}$.  It should be noticed that CVO are multivalued
functions of $z$, but the invariance of conformal fields under the
transformation 
\be 
U \ = \ e^{2 \pi i ( \ L_0 \ - \ \bar{L}_0 \ )}  
\label{spintra}
\ee
forces the conformal weights to obey the relation
$\Delta_r - \bar{\Delta}_{\bar{r}} \in {\bf Z}$ for each
representation $r$.  Moreover 
CVO, exhibiting non trivial monodromies, satisfy a braid group statistics. 
They are not uniquely determined by eq. (\ref{confields}).  On the
contrary, the gauge freedom in the definition of CVO reflects itself in the
nature of conformal fields as invariant tensors of a {\it quantum} symmetry
\cite{quantgr}.

We are in general interested in the calculation of $n$-point correlation
functions of conformal fields on a genus $g$ Riemann surface.  These depend
on the positions $( z_i, \bar{z}_i ), i = 1,...,n$ of the fields and of the $3
g - 3$ complex moduli of the Riemann surface. 
Due to the factorization of eq. (\ref{confields}), $n$-point correlation
functions are sesquilinear forms on the moduli space of the punctured Riemann
surface \cite{fs} \be
W_n \ = \ \sum_{I , \bar{I}} \ g_{I \, \bar{I}} \ {\cal F}_I \ {\bar{\cal
F}}_{\bar{I}}  \qquad . 
\label{corrfunct}
\ee
The analytic blocks $\cal F$ and $\bar{\cal F}$ are correlators of CVO and,
as such, have non-trivial monodromy and modular properties, even if the $W_n$
is single valued and modular invariant.  For instance, in minimal models they
correspond to the conformal blocks of BPZ \cite{bpz} for $g=0$, $n = 4$,
while for $g=1$, $n=0$ they are characters of the algebra $\cal
A$.  Another basic feature of RCFT is that correlators of {\it primary} CVO
are a basis of solutions of PDE's obtained from conformal Ward identities with
the use of the null vector method \cite{bpz}.  In principle, the knowledge of
the chiral  observable algebra allows us to construct the chiral correlators by
sewing three point functions on the sphere \cite{fs} \cite{vafa}
\cite{sonoda}, characterized by the OPE coefficients.    Alternatively, we can
factorize an $n$-point correlator, by degenerating in some channels the moduli
of punctured Riemann surface and using the OPE's.  This factorization
procedure is not unique and gives rise to the so called {\it sewing
constraints}.  In fact, analytic blocks are connected to each other by matrices
that represent the action of the Braid Group on the external punctures (B
matrices), of the duality transformations (F matrices) and of the modular
group generators (T and S in the genus-one case) \cite{moosei}.  It is
possible to demonstrate that only two independent sewing constraints occur in
closed CFT  \cite{sonoda} \cite{moosei}.  
They are the crossing symmetry of the four point function on the
sphere and the invariance under the ``cutting'' along the two different
homology cycles of the one point function on the torus.  
More simply, if we limit ourselves to the study of the
(perturbative) spectrum of a model, i.e. to the one-loop partition function,
all what we need is {\it modular invariance} \cite{gso} \cite{seiwit}
\cite{carmod}.  By defining as usual the characters of the algebra ${\cal A}$ 
\be \chi_{i} ( \tau ) \ = \ Tr_{{( \ \cal{H}}_{i})} \ q^{( L_{0} \ - \ c / 24
)} \qquad , \label{charact} \ee
with $q = e^{2 \pi i \tau}$, the torus partition function reads
\be
T \ = \ \sum_{i,j} \  \chi_i ( \tau ) \ N_{ij} \ {\bar \chi}_j ( {\bar \tau}
)  
\label{torus}
\ee
and must be 
invariant under the modular group 
generated by the transformations
\be
T: \ \tau \rightarrow \tau + 1 \qquad \quad {\rm and } \quad \qquad
S: \ \tau \rightarrow - \ {1 \over \tau} \quad ,
\label{TandS}
\ee
acting on characters $ \chi_i$ via two matrices, also
denoted by $T$ and $S$.  It should be noticed that 
$N_{ij}$ are $0$ or $1$ once the ambiguity between
characters with the same $q$-expansion is resolved \cite{ambig} by carefully
extending the modular matrices $S$ and $T$ so that, if $C$ is the charge
conjugation matrix, 
\be
( S )^2 \ = \ ( S T )^3 \ = \ C \qquad .
\label{modgen}
\ee
Correspondingly, the fusion-rule coefficients $N_{i j}^k$, 
connected to the $S$ matrix via the Verlinde formula \cite{verlinde}
\be
N_{ij}^{k} \ = \ \sum_{n} \ {{S_{i}^{n} \, S_{j}^{n}\, S_{n}^{\dag \, k}} \over
S_{0}^{n}} \qquad  ,
\label{verlinde}
\ee
are also integer and count the number of independent three point functions of
primary CVO.

In order to build spectra of closed (super)string models, we have to
tensorize chiral sectors of CFT (or their extension to an N=1 superconformal
algebra) in such a way that they saturate the {\it conformal anomaly} \cite{pol}
and give rise to a modular invariant genus-one partition function consistent
with spin-statistics.  The total central charge of each chiral sector in the
light-cone gauge must then be $12$ if the sector is supersymmetric and $24$ if
it is bosonic.  There are thus three classes of interesting 
strings: bosonic with $c = 
\bar{c} = 24$, heterotic with $c = 12$ and $\bar{c} = 24$ and type II with $c
=  \bar{c} = 12$.  In particular, in $d$ transverse dimensions, $(3 d / 2)$ is
the contribution to the central charge of space-time supersymmetric 
coordinates and $(12 - 3 d / 2)$ is the one of the
``internal'' theory, while $d$ and $(24 - d)$ are the corresponding values in
the bosonic case.  

It is worth at this stage to illustrate as simple examples
the partition functions of all interesting type II superstrings in
ten dimensions ($d = 8$).  They are written in terms of characters of the
level-one so(8) representations
\ba
& &O_8 = 	{1 \over {2 \eta^4}} \biggl( \ \theta^4 \biggl[ {0 \atop 0} \biggr]  \ + \
\theta^4 \biggl[ {0 \atop 1/2} \biggr] \ \biggr)
\quad \ , \quad V_8 =	{1 \over {2 \eta^4}} \biggl( \ \theta^4 
\biggl[ {0 \atop 0} \biggr] \ - \
\theta^4 \biggl[ {0 \atop 1/2} \biggr] \ \biggr) \quad , \nonumber \\
& &S_8 = 	{1 \over {2 \eta^4}} \biggl( \ \theta^4 \biggl[ {1/2 \atop 0} \biggr]  \ + \
\theta^4 \biggl[ {1/2 \atop 1/2} \biggr] \ \biggr) \ ,
\quad C_8 =	{1 \over {2 \eta^4}} \biggl( \ \theta^4 \biggl[ {1/2 \atop 0} \biggr] \ - \
\theta^4 \biggl[ {1/2 \atop 1/2} \biggr] \ \biggr) 	\ ,
\label{soottoc} 
\ea
with $\eta$ the Dedekind function and $\theta {\alpha \brack \beta}$ the
theta functions with characteristics.  If we omit the integration over moduli
and the fixed contribution of the eight transverse bosonic coordinates, the
left-right symmetric modular invariants are \cite{gso} \cite{seiwit} \cite{dh}
\ba 
&T_{IIB} \ &= \ {| V_8 \ - \ S_8 |}^2  \qquad , \\
&T_{0A} \ &= \ {| O_8 |}^2 \ + \ {| V_8 |}^2 \ + \ S_8 \ {\bar{C}}_8 \ + \
C_8 \ {\bar{S}}_8 \qquad , \\
&T_{0B} \ &= \ {| O_8 |}^2 \ + \ {| V_8 |}^2 \ + \ {| S_8 |}^2 \ + \
{| C_8 |}^2 \qquad .
\label{summpf}
\ea
The first is the ``Type IIB'' superstring while the other two, being
tachyonic and not supersymmetric, have to be considered as toy models but are
interesting for the open-string program.  The other (not left-right
symmetric) well known model in ten dimension is the ``Type IIA'' superstring,
whose partition function results
\be
T_{IIA} \ = \ ( V_8 \ - \ S_8 ) ({ \bar{V}}_8 \ - \ { \bar{C}}_8) \qquad .
\label{duea}
\ee
\vskip 24pt
\begin{flushleft}
{\large \bf 3. \ Open and unoriented CFT and superstrings}
\end{flushleft}

The basic building blocks of open (super)strings are CFT defined on arbitrary
Riemann surfaces.  This amount to take into account, in addition to theories
on closed orientable Riemann surfaces, also theories propagating on surfaces
with boundaries and/or crosscaps, three crosscaps being equivalent to one
handle and one crosscap \cite{aaa}.  This extended geometrical framework makes
necessary the inclusion of additional data in order to properly define the 
CFT.  First of all, the introduction of boundary (or open)
fields $\psi_{i}^{a b}$ beside the bulk conformal fields (\ref{confields}) is
required \cite{cardy89} \cite{cardycon}.  $\psi_{i}^{a b}$ lives exactly on
boundaries and its insertion changes type $a$ boundary conditions into type
$b$ ones.  The crucial observation is that each Riemann surface with holes
and/or crosscaps admits a closed orientable double cover and can be thus
defined as the quotient of the double cover by an (anticonformal) involution
\cite{ss}.  As a consequence, the two holomorphic and antiholomorphic chiral
algebras are no longer independent.  On the one hand, this implies the existence
of a chiral algebra that contains the ``diagonal''
combination of holomorphic and antiholomorphic Virasoro algebras
\cite{cardycon}.  On the other hand, $n$-point correlation functions in the 
presence of holes and/or crosscaps become real linear rather than sesquilinear
combinations of analytic blocks.  The presence of open fields makes also
necessary the introduction of two more OPE's \cite{lew} \cite{carlew}. One is
the product of two boundary operators, expressible in terms of the three point
functions on the disk $C_{i j k}^{a b c}$.  The other concerns the behaviour of
a bulk field $\phi_{k, \bar{k}}$ approaching a boundary.  What happens is that
the bulk field ``collides'' with its image and can be expanded in terms of only
boundary operators.  The corresponding OPE coefficients $C_{(k,\bar{k}) i}^a$
are related to the amplitude on the disk with an ``open'' and a ``closed''
puncture.  Moreover, the algebra of boundary fields acts on the same
Hilbert space as the (chiral part of) the algebra of the bulk fields.  As a
result, the field normalization is in general diverse and other coefficients
are needed to complete the definition of the CFT.  To be precise,
there is the necessity of coefficients $\alpha_{i}^{ab}$ \cite{lew} 
for taking into account
the normalization of the two-point function of boundary fields and coefficients
$\Gamma_k$ \cite{fps} responsible for the normalization of the one-point
function of bulk fields $\phi_{( k, \bar{k})}$ in front of a crosscap.  This
one-point function is essentially a chiral two-point function on the sphere of
a CVO with its image under the involution, and can be calculated as a vacuum
amplitude with an insertion of the ``crosscap operator'' \cite{pss2}
\be
{\hat C} \ = \ \sum_k \ \Gamma_k \ | \Delta_k > < {\bar \Delta}_k | \qquad ,
\label{crosscap}
\ee
where $| \Delta_k >$ corresponds to the primary CVO \ $V_k$.  In particular
\be
{< \phi_{1, \bar 1} \ .. \ \phi_{n, \bar n} >}_C
\ = \ {< {\hat C} \phi_{1, \bar 1} \ .. \ \phi_{n, \bar
n} >}_0
\quad , \label{npoint1}
\ee
and, by eq. (\ref{crosscap}), the last expression is a chiral $2n$-point
correlation function. It should be noticed that, due to the non-trivial
topology of the crosscap, a bulk field crossing a crosscap can emerge
identical or opposite to its image .  Denoting this sign with $\varepsilon$
and indicating with $X$ any polynomial in the fields, one finds
\be
{< \phi_{k, \bar k} (z_k, {\bar z}_k ) \ X >}_C \ = \
\varepsilon_{( k, \bar k )} \
{< \phi_{\bar k, k} ({\bar z}_k, z_k ) \ X >}_C \quad ,
\label{epsilo}
\ee
where the image field is obtained from the original field by interchanging
holomorphic and antiholomorphic parts.  
Finally, the normalization coefficients $B_{k}^{a}$ of the one-point
function of $\phi_{( k, \bar{k})}$ in front of a boundary, also necessary, are
a subset of the bulk-boundary OPE coefficients, namely 
$B_{k}^{a} = C_{(k,\bar{k}) \bf{1}}^a \alpha_{\bf 1}^{aa}$.  

In principle again, we are able to build every correlation function of open and
bulk fields on arbitrary Riemann surfaces by sewing the three building blocks
corresponding to the OPE previously introduced and taking into account the
normalization coefficients.  Making the ``sewing procedure'' unambiguous
produces again a finite number of sewing constraints.  
The analisys of all
sewing constraints is beyond the scope of this talk.  The interested reader can
refer to the original literature \cite{sonoda} \cite{moosei} \cite{lew} 
\cite{fps} \cite{pss2}.  Rather, it is worth discussing one of
them, the {\it ``crosscap constraint''} introduced in refs.
\cite{fps} \cite{pss2}, that determines the whole spectrum in the
non-orientable closed sector and can be exactly solved for large classes of
models like, for instance, the infinite series of A-D-E 
minimal models and $SU(2)$ WZW models of CIZ \cite{ciz}.  As
previously mentioned, the crosscap can be defined by identifying, via an
anticonformal involution, opposite points on the Riemann sphere.  As a result,
only an SU(2) subgroup of the (global) conformal group SL(2,C) ``descends'' to
the crosscap.  Actually, this residual symmetry encoded in the one-point
function on the crosscap already constraints the values of $\Gamma_k$. 
Indeed, it is easy to show that the one-point function of a field coincide
with the one-point function of the image: \be
 { < \phi_{k, {\bar k}} (z, {\bar z} )  >}_C \ = \ \Gamma_{k} \
\delta_{\Delta_k, {\bar \Delta}_k} \ < 0 | V_{k}(z) \ V_{\bar k}({\bar z}) | 0
>\ = \ {< \phi_{{\bar k}, k} ({\bar z}, z )  >}_C \quad .
\label{onepoint} 
\ee 
As a consequence, $\Gamma_k$ is vanishing for fields with non-zero spin
($\Delta_k \neq {\bar \Delta}_k)$ or with $\varepsilon = -1$, with
$\varepsilon$ the phase in eq. (\ref{epsilo}).  The two-point function in
front of a crosscap is the relevant amplitude for the ``crosscap
constraint''.  In deriving it, there is in fact an ambiguity in
inserting the crosscap state of eq. (\ref{crosscap}).  Indeed, $\hat{C}$ can
be put between the punctures $1$ and $2$ and the two images $\bar{1}$ and
$\bar{2}$, else it can separate punctures $1$ and $\bar 2$ from the remaining
$\bar 1$ and $2$ (Fig 1).  \vskip 12pt 
\input epsf \centerline{ \epsfbox{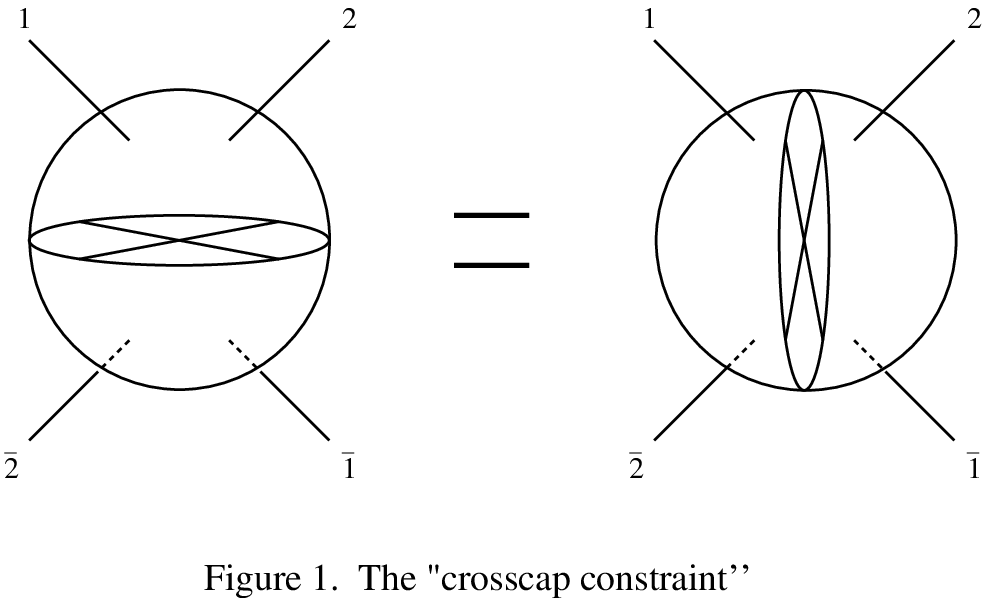}}
\vskip 12pt
Technically, using the fusion matrices, we can relate the two-point function
\be
< \phi_{(1, \bar 1)} (z_1, {\bar z}_1 )  \ \phi_{(2, \bar 2)} (z_2,
{\bar z}_2 )  >_C  \ = \  \sum_k  \Gamma_k  \  C_{( 1, \bar 1 )( 2,
\bar 2 )}^{( k, k )} \ S_{k}(z_1, z_2,{\bar z}_1,{\bar z}_2 )
\label{twopoint1} 
\ee
to the two point function
\be
< \phi_{(\bar 1, 1)} ({\bar z}_1, z_1)  \ \phi_{(2, \bar 2)} (z_2,
{\bar z}_2 )  >_C  \ = \  \sum_k  \Gamma_k  \  C_{( \bar 1, 1 )( 2,
\bar 2 )}^{( k, k )} \ S_{k}({\bar z}_1,z_1, z_2,{\bar z}_2 ) \quad ,
\label{twopoint2} 
\ee
where $S_k$ is the $s$-channel conformal block with a $k$ field in the
intermediate state and the $C$'s are the closed OPE coefficients.  Indeed, the
following relation holds 
\be
S_{k}({\bar z}_1, z_2, z_1,{\bar z}_2 ) \ = \ {( - 1 )}^{{\Delta}_1 - {\bar
\Delta}_1
+ {\Delta}_2 - {\bar \Delta}_2} \ \sum_n \ F_{kn} ( 1,2,{\bar 1},{\bar 2} ) \
S_{n}(z_1, z_2, {\bar z}_1,{\bar z}_2 ) 
\label{sduality} \quad ,
\ee
where the phase comes from a braiding $B_1 (B_3)^{-1}$ and $F$'s are the fusion
matrices.  Inserting this expression in eq. (\ref{twopoint2}) and using 
eq.(\ref{epsilo}) we obtain the ``crosscap constraint'', a linear
relation between the one-point coefficients $\Gamma_k$
\be
\varepsilon_{( 1 ,{\bar 1} )} \ {( - 1 )}^{{\Delta}_1 - {\bar
\Delta}_1
+ {\Delta}_2 - {\bar \Delta}_2} \ \Gamma_n \ C_{( 1, \bar 1)( 2,
\bar 2 )} ^{( n, n )} \ = \ \sum_k \ \Gamma_k  \ C_{( \bar 1 , 1
)(2, \bar 2 )} ^{( k, k )} F_{kn} ( 1,2,{\bar 1},{\bar
2} ) \quad . 
\label{capconstr} 
\ee

Eq. (\ref{capconstr}) plays a fundamental role also in discussing rules to
construct perturbative 
spectra of open and unoriented (super)strings, fully encoded 
in the one-loop partition function.  To tackle the whole construction from this
equivalent point of view, two preliminary observations are needed.  First, it
has long been known that a theory of only boundary operators
is inconsistent, the reason being that bulk fields are always present in the
intermediate states of non-planar open diagrams \cite{lovel}.  Second,
enclosing the non orientable contribution, i.e. allowing a ``twist''
\cite{twist} of strings, is demanded, as we will see, by the structure of
ultraviolet divergencies and of anomaly cancellations.
Four contributions enter the one-loop partition function.  The
starting one is the torus contribution of eq. (\ref{torus}) that encodes the
spectrum of the closed oriented ``parent'' model.  In order to construct a
class of ``open descendants'', we have to project the closed spectrum to a
non-orientable one.  This is obtained adding to the (halved) torus the 
(halved) Klein
bottle amplitude.  Then we have to add the two (halved) open contributions,
the annulus and M\"obius strip amplitudes, that describe the open unoriented
spectrum.  The construction is very reminiscent of what happens
in the $Z_2$ orbifolds \cite{dhvw}, where the closed spectrum is projected in a
$Z_2$-invariant way and ``twisted'' sectors corresponding to strings closed
only on the orbifold are added and projected.  Since open strings are in some
sense closed on the double cover, the orbifold should be thought in the
parameter space rather than in the target space \cite{cargese} 
\cite{hor1}.  The open
states are then analog to ``twisted sectors'', while the role of $Z_2$ group is
played by the twist that interchanges left (holomorphic) and right
(antiholomorphic) sector.  This is the reason why {\it only left-right
symmetric ``parent'' closed models can admit a class of open descendants}. 
The amplitude in eq. (\ref{torus}) must thus refer to models with
identical holomorphic and antiholomorphic sectors (the Type IIA
superstring, for instance, does not admit open descendants with 10-D Lorentz 
symmetry).  

To understand
better the consistency conditions, let us take a closer look at the three
additional amplitudes.  The (direct channel) Klein bottle amplitude, that
projects the closed spectrum of eq. (\ref{torus}), has the general form
\be
 K \ = \ {1 \over 2} \ \sum_i \ \chi_i \ K_i \qquad 
\label{kleindir}
\ee
with $| K_i | = N_{ii}$.  Actually, more choices are possible according 
to the signs of $K_i$, that determine the
(anti)symmetrization properties of the ``$i$'' sector.  The (anti)symmetry
of Verma modules must be compatible with the fusion rules. 
For instance, the fusion of two states in antisymmetrized sectors must
produce symmetric states.  In particular, the available
choices corresponds to the $Z_2$-automorphism of the fusion algebra compatible
with the torus partition function (see also ref. \cite{hor2} for an 
alternative derivation based on Chern-Simons theory on orbifolds).  Performing
an $S$ modular transformation on
$K$, we turn to the ``transverse channel'' Klein bottle amplitude that
describes the propagation of the closed spectrum between two crosscaps
states (Fig. 2).
\vskip 12pt 
\input epsf \centerline{ \epsfbox{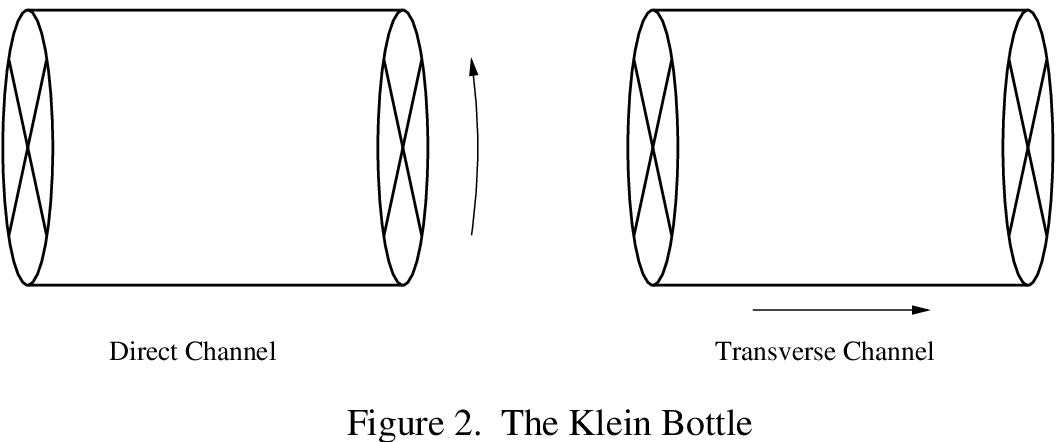}}
\vskip 12pt
It has the general form 
\be
\tilde K \ = \ {1 \over 2} \ \sum_i \ \chi_i \left( \Gamma_i \right)^2 \qquad
,
\label{kleintra}
\ee
where $\Gamma_i$ are the one-point coefficients of eq
(\ref{crosscap}).  As previously stated, they are completely determined by
solving the ``crosscap constraint''.  Notice that coefficients in $\tilde
K$ are perfect squares, while the gammas are directly related to the
phase choices in $K$ and are consistent with the fusion rules.

Let us now come to the description of the open spectrum, the genuine new
ingredient of these models.  It has been known for a long time that open-string
ends carry (Chan-Paton) charges \cite{cp} \cite{quarks} 
that manifest themselves, at the
level of partition functions, as multiplicities to be associated to the annulus
and M\"obius amplitudes.  After all, in conventional orbifolds multiplicities
are associated to fixed points, and boundaries are fixed under the involution. 
Let us start from the transverse annulus amplitude that describes the closed
spectrum flowing between two boundary states (Fig. 3).
\vskip 12pt 
\input epsf \centerline{ \epsfbox{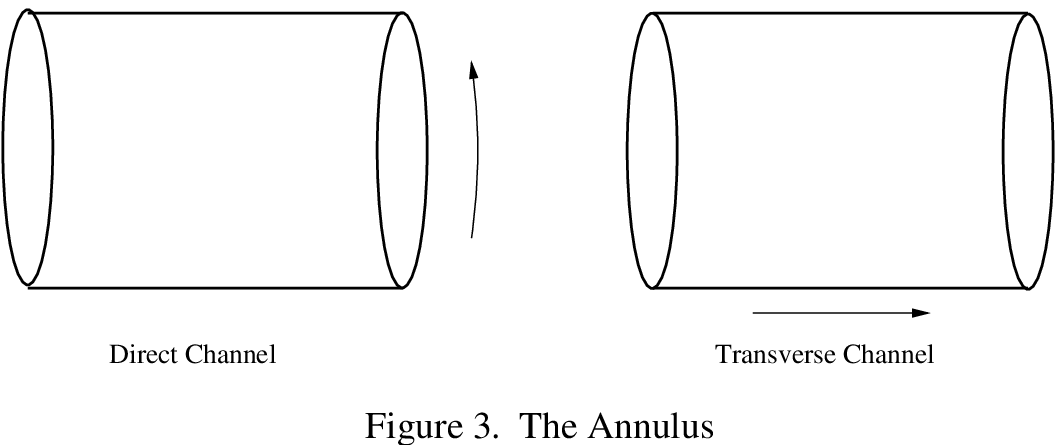}}
\vskip 12pt
Due to the mirror-like
properties of the boundary, {\it only fields which are paired with the
conjugates in the closed GSO projection of eq.} (\ref{torus}) {\it can appear
in the tranverse annulus} \cite{bs}.
Technically, they are the only states with
non-vanishing $B_{k}^a$ coefficients.  If we introduce the Chan-Paton
multiplicities $n^a$, the transverse annulus amplitude can be written
\be
\tilde{A} \ = \ {\frac{1}{2}} \ \sum_{k} \, \chi_k \ \left( \sum_{a} \,
B_{k}^{a} \, n^a \, \right)^2  \qquad  ,
\label{aneltr}
\ee
where the perfect squares indicate, as expected, the presence of two
boundaries in the annulus.  Performing as in the Klein bottle amplitude the $S$
modular transformation exposes the direct-channel annulus amplitude, that reads
\be
A \ = \ {\frac{1}{2}} \sum_{k, a, b} \ A_{a b}^k \ n^a \ n^b \ \chi_k \qquad .
\label{aneldi}
\ee
The non negative integer coefficients $A_{ab}^k$ are very important, because
they determine the open spectrum or, equivalently, classify the set of
conformally invariant boundary conditions.  The interpretation of eq.
(\ref{aneldi}) is that open states of the $k$ sector with charges $n^a$ and
$n^b$ and multiplicities corresponding to $A_{ab}^k$ can exist if $A_{ab}^k$
itself is non-vanishing.  In the {\it diagonal} RCFT, the flow of the $k$-th
sector along a strip with boundary condition $a$ and $b$ is governed by the
fusion-rule coefficients \cite{cardy89}.  This implies that $A_{ab}^k =
N_{ab}^k$ (diagonal ansatz), and using eq. (\ref{verlinde}) allows to express
the $B_{k}^a$ in terms of entries of the modular matrix $S$ \cite{cardy89}
\cite{carlew}: 
\be
B_{k}^a \ = \ \frac{S_{k}^a}{\sqrt{S_{k}^1}} \qquad .
\label{bi}
\ee
In the more complicated non-diagonal models, $B_{k}^a$ are determined by
solving suitable sewing constraints and $A_{ab}^k$ are no-longer the
fusion-rule coefficients \cite{pss2}.  Rather, they satisfy completeness
relations and can manifest the presence of extended boundary operator algebras,
as compared to the diagonal case \cite{pss3}.  Finally, the direct channel
M\"obius amplitude (anti)symmetrizes the open spectrum, and has the form
\be
M \ = \ \pm \, {1 \over 2} \, \sum_{k,a} \, M_{a}^k \, n^a \, {\hat{\chi}}_k 
\qquad ,
\label{moebdi}
\ee
where $M_{a}^k = A_{a a}^k$ (mod 2) and $\hat{\chi}_k$ are the suitable
characters that count states on the M\"obius strip, whose modulus is not
purely imaginary.  The overall sign is free in RCFT, but is crucial in
critical open-string models where it determines the type of gauge (Chan-Paton)
groups.  The transverse channel M\"obius contributions that represent closed
states flowing between a hole and a crosscap (Fig. 4) 
\vskip 12pt 
\input epsf \centerline{ \epsfbox{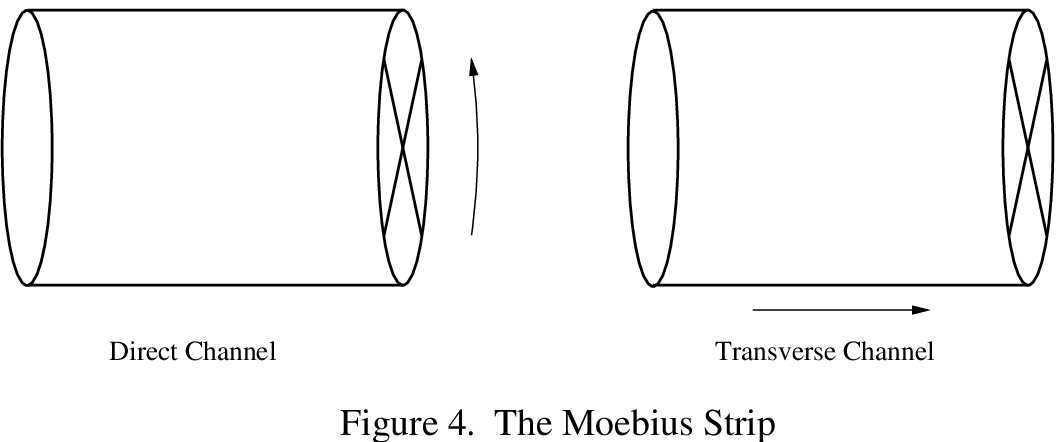}}
\vskip 12pt
are,
as intuitively expected, geometric means of those in $\tilde{A}$ and
$\tilde{K}$: \be
\tilde{M} \ = \ \pm \, \sum_{k} \, \left( \, \sum_{a} \,
B_{k}^a \, n^a \, \right) \ \Gamma_k \
\hat{\chi}_k  \qquad . \label{moebtra}
\ee
In order to obtain $M$ from $\tilde{M}$ and thus determine the signs in
$M_{k}^a$ the matrix $P=T^{1/2} \ S \ T^2 \ S \ T^{1/2}$ must be used that
implements on the basis of ``hatted'' characters the modular transformation
${i \tau  + 1 \over 2} \rightarrow {i   + \tau \over 2 \tau }$.  The
compatibility between $M$ and $\tilde{M}$ is the last constraint.  It
should be stressed that given a left-right symmetric closed oriented model,
the algorithm allows to build a class of consistent open descendants.  In
particular, open amplitudes automatically satisfy planar duality and
factorization properties \cite{bps} \cite{pss1}, as we shall see in an
example in section 4.  Two final observation are in order: 
first, open states can
be, in some sectors, oriented.  This reflects itself in the presence of complex
Chan-Paton charges \cite{bs}.  Second, in critical (super)string models the
tranverse amplitudes $\tilde{K}$, $\tilde{A}$ and $\tilde{M}$ exhibit the flow
of massless scalars that can acquire a VEV (tadpoles) and in principle have to
be eliminated from the spectrum.  This requires the cooperative action of all
transverse channel amplitudes and justifies the introduction of
the non-orientable contributions \cite{quarks} \cite{tadpoles}.  While the
presence of tadpoles of ``physical'' scalars signal a vacuum instability that
could hide a Higgs-like phenomenon, tadpoles of ``unphysical'' scalars produce
inconsistencies and must be eliminated.  It is possible to show that the
cancellation of tadpoles of ``unphysical'' scalars is equivalent to the
cancellation of all gauge and gravitational anomalies in the low energy
effective field theory \cite{pcay}.

\vskip 24pt
\begin{flushleft}
{\large \bf 4. \ Examples}
\end{flushleft}

In order to appreciate the algorithm described in the preceding section, it is
useful to analyze some concrete examples.  In particular, let us start by
deriving the open descendants of the Ising model, the simplest model in the
infinite series of minimal $A-A$ modular invariants, together with the 
descendants of $A_3$ model, the simplest non trivial model in the $A$-series of
$SU(2)$ WZW modular invariants \cite{pss1}.  Both are diagonal
and the open descendants are based, as said, on the diagonal ansatz.  However,
the presence of the $SU(2)$ Kac-Moody algebra in the $A_3$ case reflects itself
in a number of subtleties that are worthy of a detailed discussion and are
common to all models in the two infinite series.

The Ising and $A_3$ models on the torus share formally the same partition
function 
\be
T= |\chi_1|^2 \ + \ |\chi_2|^2 \ + \ |\chi_3|^2	\qquad ,
\label{atretoro}
\ee
provided one identifies $\chi_1$, $\chi_2$ and $\chi_3$ with the characters of
identity, spin and energy for Ising, and with $\chi_{2I+1}$, with $I$ the
isospin, for $A_3$.  The $S$ matrix is identical as well,
\be
S \, = \, {1 \over 2}
\pmatrix{1&\sqrt{2}&1 \cr \sqrt{2}&0&-\sqrt{2} \cr 1&-\sqrt{2}&1} \quad ,
\label{sk2}
\ee
but open descendants must be different because, due to the fact that 
the conformal
dimension of $\chi_2$ is $3/16$ while that of the Ising spin is $1/16$,
their $P$ matrices are 
\be
P_{(Is)} \ = \ \pmatrix{c&0&s \cr 0&1&0 \cr s&0&-c} \qquad, \qquad
P_{(A_3)} \ = \ \pmatrix{s&0&c \cr 0&1&0 \cr c&0&-s} \qquad, 
\label{pmatrices}
\ee 
with $s=\sin(\pi/8)$ and $c=\cos(\pi/8)$.  There are two possible Klein bottle
projections, in correspondence with the only possible automorphism of the fusion
rules, the $Z_2$ center of $SU(2)$ in $A_3$ and the spin reversal in the Ising
model: 
\be
K_1 \ = \ {1 \over 2} \ \left( \, \chi_1 \ + \ \chi_2 \ + \ \chi_3 \, \right) 
\qquad , \qquad K_2 \ = \ {1 \over 2} \ \left( \, \chi_1 \ - \ \chi_2
\ + \ \chi_3 \, \right)  \qquad .
\label{kleins}
\ee
It should be noticed, however, that $K_1$ corresponds to having the Ising
spin symmetric in front of the crosscap (i.e. the $\varepsilon$ of
eq. (\ref{epsilo}) equal to $+1$) and $K_2$ to having the Ising
spin antisymmetric ($\varepsilon =
-1$).  On the contrary, $K_1$ corresponds to having the isospin $1/2$ field of
$A_3$ antisymmetric ($\varepsilon = -1$) in front of the crosscap and $K_2$
to having it 
symmetric ($\varepsilon = +1$).  The open spectrum feels the signs, because
they enter the ``crosscap constraint'' and thus the coefficients $\Gamma_k$,
that appear in $\tilde{K}$ and in $\tilde{M}$.  As a
consequence, we have four different classes of open descendants.  $K_1$ leads
to descendants with real Chan-Paton charges for Ising
\ba
&A1_{(Is)} \,  &= \, 
\biggl(  {{n_0^2 + n_{1/2}^2 + n_{1/16}^2}\over{2}} \biggr) \chi_0 \ + \
n_{1/16} ( n_0 + n_{1/2} ) \chi_{1/16} \ + \ \biggl({{n_{1/16}^2} \over 2} +
n_0 n_{1/2} \biggr) \chi_{1/2} \, , \nonumber \\ 
&M1_{(Is)} \, &= \ \pm \ \biggl[ \ {{n_0 + n_{1/16} + n_{1/2}}\over{2}} \
\hat{\chi}_0 \ + \ {{n_{1/16}} \over 2} \ \hat{\chi}_{1/2}
\ \biggr]
\qquad ,
\label{isireal}
\ea  
and with complex Chan-Paton charges for $A_3$
\ba
&A1_{(A_3)} \, &= \ 
\biggl(  {{n_2^2}\over{2}} + m \bar{m} \biggr) \ \chi_1 \ + \
n_2 \bigl( m + \bar{m} \bigr) \ \chi_2 \ + \
{{n_2^2 + m^2 +\bar{m}^2} \over 2} \ \chi_3 \, , \nonumber \\
&M1_{(A_3)} \, &= \,\pm \ \biggl[ \ {{n_2}\over{2}} \ \hat{\chi}_1 \ + \
{{n_2 + m +\bar{m}} \over 2} \ \hat{\chi}_3 \ \biggr] \qquad .
\label{atrecom} \ea
The opposite is true for $K_2$, that leads to descendants
with complex Chan-Paton charges for Ising
\ba
&A2_{(Is)} \, &= \ \biggl(  {{n_{1/16}^2}\over{2}} + m \bar{m}
\biggr) \ \chi_0 \ + \ n_{1/16} \bigl(  m + \bar{m} \bigr) \
\chi_{1/16} \ + \ {{n_{1/16}^2 + m^2 +\bar{m}^2} \over 2} \ \chi_{1/2}
\, , \nonumber \\ &M2_{(Is)} \, &= \,\pm \
\biggl[{{n_{1/16}}\over{2}} \ \hat{\chi}_0 \ + \ {{ m +\bar{m} -
n_{1/16}} \over 2}  \ \hat{\chi}_{1/2} \ \biggr] \qquad ,  
\label{isicom} 
\ea
and to descendants with real Chan-Paton charges for $A_3$
\ba
&A2_{(A_3)} \, &= \ \biggl(  {{n_1^2 + n_2^2 + n_3^2}\over{2}} \biggr)
\chi_1 \ + \ n_2 ( n_1 + n_3 ) \chi_2 \ + \ \biggl({{n_2^2} \over 2} +
n_1 n_3 \biggr) \chi_3 \, , \nonumber \\
&M2_{(A_3)} \, &= \,\pm \ \biggl[ \ {{n_1 - n_2 + n_3}\over{2}} \
\hat{\chi}_1 \ + \ {{n_2} \over 2} \ \hat{\chi}_3 \ \biggr] \qquad . 
\label{atrerea} 
\ea
This structure repeats itself for the whole
$A-A$ series of minimal models and for the whole $A$-series of $
SU(2)$ WZW models.  Given a modular invariant
torus partition function, there exist two classes of open descendants with
real or complex charges.  For instance, the modular invariant of the $A$
series at level $k$, in the same notation of eq. (\ref{atretoro}), is the
diagonal one 
\be
T \ = \ \sum_{a=1}^{k+1} |\chi_a|^2 \qquad ,
\label{torusa}
\ee
and the Klein bottle projection leading to all real charges is
\be
K \ = \ {1 \over 2} \ \sum_{a=1}^{k+1} (-1)^{(a-1)} \chi_a  \qquad  .
\label{kleinar}
\ee
The annulus amplitude is directly obtainable by the diagonal ansatz, while the
M\"obius amplitude exhibits signs reflecting the underlying current algebra
\ba
&A \ &= \ {1 \over 2} \ \sum_{a,b,c} \ N_{ab}^{c} \ n^{a} \
n^{b} \ \chi_{c} \quad, \nonumber \\
&M \ &= \ \pm {1 \over 2} \ \sum_{a,b} \  (-1)^{b-1} \ (-1)^{{a-1} \over
2} \ N_{bb}^{a} \ n^{b} \ \hat{\chi}_{a} \qquad .
\label{opena}
\ea
In particular, as discussed for $A_3$, the phase $(-1)^{{a-1} \over 2}$ is
due to the behaviour of fields in front of the crosscap and the phase
$(-1)^{b-1}$ changes the type of charges in a way corresponding to the isospin.
Moreover, the coefficients $\Gamma_k$ can nicely be expressed in terms of $S$
and $P$ matrices
\be
\Gamma_k \ = \ {{P_{1k}} \over {\sqrt{S_{1k}}}} \qquad , 
\label{gamma}
\ee
and allow direct channel $K$ and $M$ amplitudes to be written using some
components of the integer-valued tensor \cite{pss1} 
\be
Y_{abc} \ = \ \sum_{d} \ {{S_{ad} \ P_{bd} \ {P^\dagger}_{cd}} \over
{S_{1d}}} \qquad . 
\label{ymat}
\ee 
It is also interesting to briefly mention how {\it relative} signs in the
non-orientable contributions are connected to the action of ``twist'' 
respecting factorization and planar duality of amplitudes
\cite{bps} \cite{pss1}.  To this end, it is sufficient to consider the Ising
model.   
\vskip 12pt
\input epsf \centerline{ \epsfbox{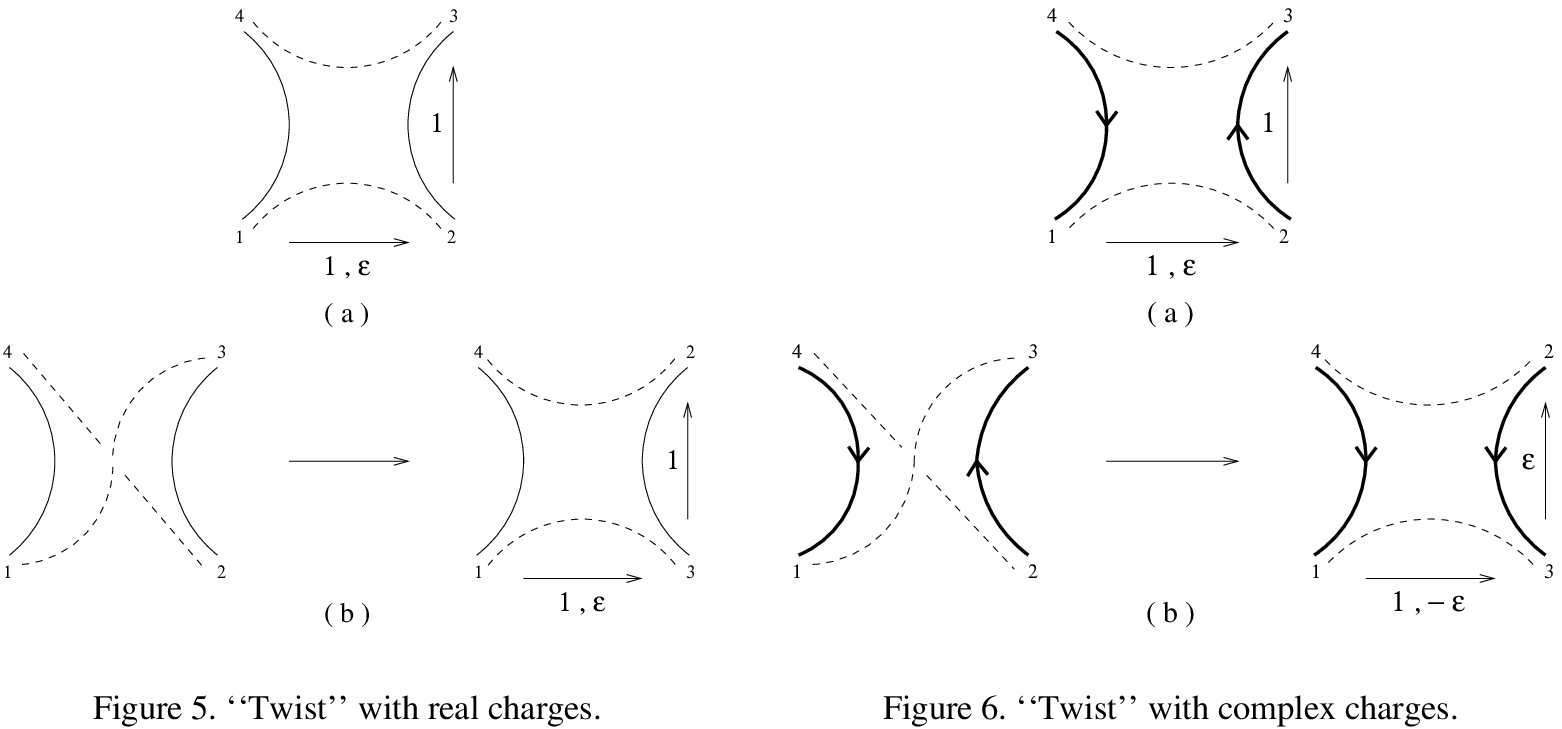}}
\vskip 12pt
Referring to figure 5, let us first analyze the four-spin amplitude on
the disk for the model with real charges.  Due to the fusion rules, indicating
with dashed lines $n_{1/16}$ charges and with continuous lines $n_0$ charges,
only the identity flows in the $s$-channel, while identity and energy flow
in the $u$-channel (Fig. 5a).  If we take into account the ``twisted''
$u$-channel amplitude, we can see that it is related, by an operation
of ``unfolding'', to an amplitude in which the identity flows in the
$t$-channel and again identity and energy flow in the $u$-channel with the
same sign.  This means that identity and energy with a pair of $n_{1/16}$
charges ``twist'' in the same way, as demanded by $A1_{(Is)}$ and
$M1_{(Is)}$ in eq. (\ref{isireal}).  Let us then consider the same four-spin
amplitude, but referred to the model with complex charges (Fig. 6).  If
dashed lines indicate again $n_{1/16}$ charges, we have now to introduce
arrows associated to complex charges $m$ and $\bar m$.  An inspection of the
partition function reveals that in the $s$-channel only the identity with
charges $m$ and $\bar m$ can flow (fig 6a).  By fusion, again the $u$ channel
``sees'' identity and energy with a ``pair'' of $n_{1/16}$ charges. 
However, $u$-channel ``twist'' is now no longer ``ineffective'' 
on charges,  since the unfolding
inverts the relative orientation of arrows.  As a result, the twisted 
amplitude admits only the energy flowing in its $t$-channel and consequently
identity and energy in its $u$-channel, but with {\it opposite} signs,
as demanded by $A2_{(Is)}$ and $M2_{(Is)}$ in eq. (\ref{isicom}).  The
opposite sign in the behaviour under twist of $\chi_1$ and $\chi_3$ in the
$A_3$ model with respect to the identity and energy of Ising may be traced 
to the opposite sign in the behaviour of the corresponding
conformal blocks under the action of the fusion matrix $F$ \cite{pss1}.

Open descendants of modular invariants of $D$ and $E$
type can be constructed as well.  In particular, $E_{even}$ and $D_{even}$
models exhibit an extended algebra and become (quasi)diagonal after the
resolution of ambiguity.  This introduces factors of two in some fusion rules,
corresponding to the possibility of having some multiple families with same
charges and also multiple three point open functions.  The off-diagonal models
$E_{7}$ and $D_{odd}$ are less conventional because, as said in section 3,
they are not directly based on the fusion rules, and a number of subtleties
arise from boundary operators and related sewing constraints \cite{pss3}.

Let us conclude by displaying some examples concerning open (super)strings.
There is only one open descendant of the Type II superstring in $D=10$,
the long known Type I SO(32) superstring \cite{gs}.  Indeed, the closed
spectrum exhibits a single chiral ``supercharacter'' corresponding to the GSO
projection, namely $V_{8} - S_{8}$.  It flows also on the annulus and is
modular invariant by itself.  Only one Chan-Paton charge is present and the
direct channel contributions are 
\ba K_{IIB} \ &=& \ {1 \over 2} \, \left(
\, V_{8} \ - \ S_{8} \, \right)  \qquad , \\ A_{IIB} \ &=& \ {1 \over 2} \ n^2
\ \left( \, V_{8} \ - \ S_{8} \, \right) \qquad , \\  
M_{IIB} \ &=& \ \pm \, {1 \over 2} \ n \ \left( \, V_{8} \ - \
S_{8} \, \right)  \qquad .
\label{dueB}
\ea
Performing the modular transformations to turn the amplitudes into the
transverse channel, crucial factors of two arise from the (omitted) modular
measure due to the fact that a comparison of different surfaces is possible
only in terms of the double cover \cite{cargese} \cite{ps}.  To be precise,
$\tilde{K}$ gains a factor $2^5$ while $\tilde{A}$ gains a factor $2^{-5}$. 
The cancellation of the ``unphysical'' massless 
scalar of the R-R sector forces the
M\"obius overall sign to be negative and the coefficient in front of $S_{8}$
to be zero: \be
{2^5 \over 2} \ + \ {n^2 \, 2^{- 5} \over 2} \ - \ n \ = \ 0 \qquad .
\label{tadpRR}
\ee
The resulting model is precisely the type I SO(32) superstring of Green
and Schwarz.  There exists also a bosonic analog of this model with gauge
group SO(8192) \cite{quarks} \cite{tadpoles}. 

Open descendants of Type $0$ models are very instructive since they
exhibit many aspects of the general construction \cite{bs}.  Let us
concentrate on the type $0B$ model, in which four Chan-Paton charge sectors are
present due to the (self)conjugation properties of the characters
(\ref{soottoc}).  Notice that only two charges corresponding to the two
characters flowing in $\tilde{A}$ would enter the Type $0A$ descendants.  There
are three possible inequivalent choices of Klein bottle projection compatible
with fusion rules and positivity of the transverse channel:
\ba
K_{0B} \ = \ {1 \over 2} \ \left( \, O_{8} \, + \, V_{8} \, - \, S_{8} \, -
\, C_{8} \right)  \qquad , \label{keiob} \\
{K'}_{0B} \ = \ {1 \over 2} \ \left( \, O_{8} \, +
\, V_{8} \, + \, S_{8} \, + \, C_{8} \right)  \qquad , \label{keipriob} \\
{K''}_{0B} \ = \ {1 \over 2} \ \left( \, V_{8} \, - \,
O_{8} \, + \, S_{8} \, - \, C_{8} \right)  \qquad .
\label{keisecob}
\ea
The first one is referred to the conventional choice of a basis in the light
cone gauge of $SO(1,9)$, where the NS-NS sector is symmetrized while the R-R
sector antisymmetrized.  The other two correspond to the basis
$(O_{8},V_{8},S_{8},C_{8})$ and $(-O_{8},V_{8},-S_{8},C_{8})$.  It should be
stressed that, while the closed spectrum surviving the ${K''}$ projection
does not contain tachyons and is chiral, both  ${K}$ and ${K'}$ leave
tachyons.  Open sectors can be
constructed directly in terms of the diagonal ansatz, using the general
prescription of eq. (\ref{aneldi}) with $A_{ab}^k = N_{ab}^k$.  However, in
the transverse channel of Klein bottle amplitudes (and, consequently, in the
transverse channel of M\"obius strip amplitudes) only one character flows:
\ba
&\tilde{K}_{0B} \ &= \ {2^6 \over 2} \ V_{8} \qquad , \label{kreal} \\
&\tilde{K'}_{0B} \ &= \ {2^6 \over 2} \ O_{8} \qquad , \label{kcomp1} \\
&\tilde{K''}_{0B} \ &= \ - {2^6 \over 2} \ C_{8} \qquad ,
\label{kcomp2} 
\ea
As a result, even in the direct channel M\"obius amplitude a single
character is present, the same as in the transverse Klein bottle amplitude,
since the $P$ matrix is diagonal in this case.  The character $V_{8}$ plays the
role of identity in the light-cone gauge of SO(1,9).  This implies that an
assignment of {\it real} Chan-Paton charges is compatible only with the
${K}_{0B}$ projection, and results in
\ba
A_{0B} \ &=& \ {{{n_o}^2 +  {n_v}^2 + {n_s}^2 +  {n_c}^2 }
\over 2} \ V_8 \ + \ ( n_o n_v +  n_s n_c ) \ O_8 \nonumber \\ 
&-&( n_o n_c \ + \ n_v n_s ) \ S_8 \ - \
( n_o n_s \ + \ n_v n_c ) \ C_8  \qquad , \label{anob}  \\
M_{0B} \ &=&  - \ {1 \over 2} \ ( n_o \ + \ n_v \ + \ n_s \ + \ n_c ) \
\hat{V}_8 \qquad . \label{moebob}
\ea
It should be noticed that the open spectrum is chiral.  An inspection of the
transverse channel reveals three tadpole conditions.  Two of them,
corresponding to R-R ``unphysical'' 
scalars, give $ n_0 = n_v $ and $ n_s = n_c $
and guarantee the cancellation of gauge and gravitational anomalies.  The third
one, ``physical'', is the tadpole relative to the dilaton.  If we cancel it,
the total Chan-Paton group dimensionality is fixed to 64, and the open sector
exhibits an $SO(n)
\otimes SO(32 - n) \otimes SO(n) \otimes SO(32 - n)$ symmetry group.

The other two projections are only compatible with complex Chan-Paton
charges.  Looking, for instance, to the more interesting model of eq.
(\ref{keisecob}), the charge assignment must be the following \cite{susy}:
\ba
{A''}_{0B} \ &=& \ ( n_1 \bar n_1 +  n_2 \bar n_2 ) \ V_8 \ + \ ( n_1 \bar n_2
+  n_2 \bar n_1 ) \ O_8 \nonumber \\  &-&( n_1 n_2 \ + \ \bar n_1 \bar
n_2 ) \ S_8 \ - \ ( {{n_1}^2 \over 2} + {{n_2}^2 \over 2} + {{\bar
n_1}^2 \over 2} +  {{\bar n_2}^2 \over 2} ) \ C_8  \qquad ,
\label{ansecob}  \\ {M''}_{0B} \ &=&  - \ {1 \over 2} \ ( n_1 \ + \
\bar n_1 \ - \ n_2 \ - \ \bar n_2 ) \ \hat{C}_8 \qquad .
\label{moesecbob} 
\ea
The gauge vector is not projected on the M\"obius strip because,
having ends carrying a ``quark-antiquark'' pair of $U(n_1) \otimes U(n_2)$,
is oriented.  The (numerical) equalities $n_1 = {\bar{n}}_1$ and $n_2 =
{\bar{n}}_2$ are necessary in order to avoid negative reflection coefficients in
front of the boundaries.  Thus, only one ``unphysical'' tadpole condition
survives, giving the constraint $n_1 - n_2 = 32$.  The dilaton tadpole cannot
be canceled in this model, but this, rather than being a problem, could open
new perspectives in the analisys of low energy models \cite{bachas}.
Large classes of open models based on superstrings built with the
free-fermion construction \cite{abk} can be consistently defined in lower
dimension \cite{bs}.  In particular, 
chiral models in $D=6$ are anomaly free thanks to
a generalized Green-Schwarz cancellation mechanism \cite{gengs}.  
Chiral four-dimensional models can also be built, but models so
far analyzed exhibit small-sized gauge groups.  
It is our opinion that this is due to the too simple
structure of rational closed models built in terms of free-fermions alone. 
Descendants of models with genuinely interacting CFT in the internal
sector, like for instance $N=2$ models \cite{gepner}, could open new
possibilities.

In conclusion, we have described how to consistently define CFT on Riemann
surfaces with holes and/or crosscaps. They constitute the basic building
blocks of open and unoriented (super)strings.  In particular, we have reviewed
the role of sewing constraints and the structure of one-loop partition
functions, giving some explicit examples of classes of open descendants of
left-right symmetric closed oriented models. 
\vskip 24pt \begin{flushleft}
{\large \bf Acknowledgments}
\end{flushleft}

I am grateful to the organizers for the kind invitation and to M. Bianchi, A.
Sagnotti and Ya.S. Stanev for the collaboration on related works.  This work
was supported in part by E.E.C. Grant CHRX-CT93-0340.

\end{document}